\documentclass[letter,twocolumn]{jpsj3} 
\usepackage{color,graphicx}

\def\runauthor{S. \textsc{Miyahara} and N. \textsc{Furukawa}}

\title{
  Nonreciprocal Directional Dichroism and Toroidalmagnons
  in Helical Magnets
}
  
\author{
  Shin \textsc{Miyahara}$^{1}$ 
  and Nobuo \textsc{Furukawa}$^{1,2}$
}

\inst{
  $^{1}$Multiferroics Project (MF), ERATO,
  Japan Science and Technology Agency (JST), \\
  c/o Department of Applied Physics, The University of Tokyo, 
  Bunkyo-ku, Tokyo 113-8656, Japan \\
  $^{2}$Department of Physics and Mathematics, Aoyama Gakuin University,
  Sagamihara 229-8558, Japan 
}
  
\abst{
  We investigate a dynamical magnetoelectric effect due to
  a magnetic resonance in helical spin structures
  through the coupling between magnetization and 
  electric polarization via a spin current mechanism. 
  We show that the magnon has both the dynamical magnetic moment 
  $\Delta M^\omega$ and the electric moment $\Delta P^\omega$
  ($\perp \Delta M^\omega$),
  i.e., a dynamical toroidal moment, 
  under external magnetic fields,
  and thus it is named the {\em toroidalmagnon}. 
  The toroidalmagnon exists in most conical spin structures
  owing to the generality of the spin current mechanism.  
  In the absorption of electromagnetic waves, 
  the toroidalmagnon excitation process generally induces 
  a nonreciprocal directional dichroism as a consequence
  of an interference of the magnetic and electric responses.
}

\kword{
  multiferroics, toroidalmagnon, nonreciprocal directional dichroism,
  spin current mechanism, frustration
}

\begin{document}
  \maketitle 
  
  Electromagnons are  magnons that accompany an electric dipole moment.  
  Electromagnon excitations are thus electro-active, i.e.,
  they are induced by the electric component of light,
  and therefore are classified 
  as dynamical magnetoelectric 
  phenomena.~\cite{pimenov06,katsura07,tokura11}
  Electromagnons are observed as resonances  in  optical spectroscopy 
  at terahertz frequencies in various materials,
  e.g., $R{\rm MnO_3}$~\cite{pimenov06,kida09,aguilar09},
  $R{\rm Mn_2O_5}$~\cite{sushkov08},
  ${\rm Ba_2Mg_2Fe_{12}O_{22}}$~\cite{kida09b}, 
  ${\rm CuFe_{1-x}Ga_xO_2}$~\cite{seki10}, and 
  ${\rm Ba_2CoGe_2O_7}$~\cite{kezsmarki11,bordacs11}.
  Their possible applications have been discussed.~\cite{tokura11,fiebig05,eerenstein06}.

  Electro-active magnetic resonance (MR) processes
  can be understood by  spin-dependent 
  electric polarizations ${\bf P}(\{{\bf S}_i\})$~\cite{tanabe65,moriya68}. 
  Through the coupling between the electric component 
  of light ${\bf E}^\omega$ and ${\bf P}(\{{\bf S}_i\})$, a spin structure 
  is modulated by ${\bf E}^\omega$. 
  When such modulation induces a single-magnon excitation in an ordered magnet,
  the excitation is called the electromagnon.
  Thus far, the electromagnons induced through
  the spin current~\cite{katsura07,tewari08,cano09},
  exchange striction~\cite{aguilar09,stenberg09,mochizuki10},
  and metal ligand hybridization mechanisms~\cite{miyahara11}
  have been proposed. 

  Although most of these mechanisms have restrictions
  with respect to lattice symmetry,
  the spin current mechanism can generally induce electric polarization
  in cycloidal spin configurations,~\cite{katsura05,arima11}
  irrespective of local lattice symmetries~\cite{kaplan08}.
  Let us consider a helical magnet with a cycloidal spin 
  configuration ${\bf h}\perp {\bf q}_0$,
  where  ${\bf q}_0$ is the propagation vector of 
  the magnetic   structure 
  and ${\bf h} = {\bf S}_i \times {\bf S}_{j} $
  gives the helicity vector. Here, ${\bf S}_{i}$ and ${\bf S}_{j}$
  are spins on adjacent sites along the ${\bf q}_0$ direction.
  The spin current mechanism induces the electric polarization
   ${\bf P} \sim ({\bf h}\times {\bf q}_0)$.
  In ref.~\citen{katsura07}, it is discussed that
  the uniform rotation of the helicity vector 
  around the $q_0$ direction 
  accompanies the dynamical fluctuation of the electric polarization; 
  thus, such magnon excitation is electro-active.
  Since such magnetic excitation is the Nambu-Goldstone mode,
  the electromagnon should ubiquitously be identified 
  in cycloidal spin systems.
  
  Another novel feature in the presence of dynamical 
  magnetoelectric effects is the
  nonreciprocal directional dichroism (NDD) where
  absorption intensity depends on the direction of 
  an electromagnetic wave propagation 
  vector~\cite{baron,landau}.
  Recently, NDD in MR has been 
  observed in ${\rm Ba_2CoGe_2O_7}$~\cite{kezsmarki11} 
  owing to the electro-active magnetic excitation process
  via the metal ligand hybridization mechanisms~\cite{arima07,miyahara11}.
  Since the observation of NDD at a magnon resonance is a direct evidence 
  of the presence of dynamical magnetoelectric effects 
  including an electromagnon process,
  the investigation of NDD is important for understanding the magnetoelectric
  properties of the system.
  The metal ligand hybridization mechanism, however, depends on
  the local lattice structure; thus, NDD due to such a mechanism
  should be observed in a limited class of materials.
  On the other hand, in the generic electromagnons 
  induced through the spin current mechanism in 
  cycloidal spin structures, NDD is not observed
  with symmetrical restriction,
  despite the fact that dynamical magnetoelectric couplings
  exist~\cite{katsura07}.

  In this Letter, we discuss that NDD should ubiquitously
  be observed if one applies a static magnetic field to helical magnets
  with spin current couplings.  
  We clarify the existence of 
  magnetic excitations  that accompany both magnetic and electric moments,
  which contribute to dynamical magnetoelectric 
  effects at their resonances.
  We discuss the existence of the dynamical toroidal moment
  in magnon excitation as the origin of NDD.

  Let us first review NDD briefly.~\cite{kezsmarki11,miyahara11}
  In the presence of dynamical magneto-electric effects, we have
  \begin{eqnarray}
    {\bf B}^\omega  & = & (\hat{\mu}^\infty 
    + \hat{\chi}^{\rm mm}) \mu_0 {\bf H}^\omega 
    + \hat{\chi}^{\rm me} \sqrt{\epsilon_0 \mu_0}  {\bf E}^\omega, 
    \label{eq:InducedB}
    \\
    {\bf D}^\omega & = & (\hat{\epsilon}^\infty 
    + \hat{\chi}^{\rm ee}) \epsilon_0 {\bf E}^\omega 
    + \hat{\chi}^{\rm em} \sqrt{\epsilon_0 \mu_0}  {\bf H}^\omega,
    \label{eq:InducedD}
  \end{eqnarray}
  where $\hat{\chi}^{\rm mm}$ and 
  $\hat{\chi}^{\rm ee}$ are dynamical
  magnetic and electric susceptibility tensors, respectively, and
  $\hat{\chi}^{\rm me}$  and  $\hat{\chi}^{\rm em}$ are 
  dynamical magnetoelectric susceptibility tensors.
  As an example,
  let us consider a linearly polarized light 
  with ${\bf E}^\omega \| x$, ${\bf H}^\omega \| y$, and 
  the propagation vector ${\bf k} \| z$.
  In the presence of off-diagonal 
  magnetoelectric susceptibilities, 
  the interferences between $E^\omega_x$ and $H^\omega_y$ 
  in ${\bf B}^\omega$ and ${\bf D}^\omega$
  given by eqs.~(\ref{eq:InducedB}) and (\ref{eq:InducedD})
  are affected by the relative sign of $E^\omega_x$ and $H^\omega_y$
  and thus by the direction of ${\bf k}$ since 
  $\omega \mu_0 {\bf H}^\omega = {\bf k} \times {\bf E}^\omega$. 
  For simplicity, we assume isotropic and constant $\epsilon^\infty$ and
  $\mu^\infty$.  From the Maxwell equation, 
  the complex refractive index $N$ is approximated as
  \begin{eqnarray}
    N (s_k) & \sim &
    \sqrt{[\epsilon_\infty+
      \chi^{\rm ee}_{xx} (\omega)]
      [\mu_\infty+\chi^{\rm mm}_{yy}  (\omega)]}
    \nonumber \\ && 
    + s_k [\chi^{{\rm me}}_{yx} (\omega)
    + \chi^{{\rm em}}_{xy} (\omega)]/2
  \end{eqnarray}
  as a function of the direction of the propagation vector $s_k = k_z/|k_z|$.
  The absorption coefficient is proportional to ${\rm Im} N(s_k)$.
  Therefore, the absorption intensity depends on the 
  light propagation direction $s_k$ and their difference 
  is proportional to
   $ {\rm Im} N(+) - {\rm Im}N(-) ={\rm Im} [\chi^{{\rm me}}_{yx}  (\omega)
  + \chi^{{\rm em}}_{xy}  (\omega)]$.
  In general,
  when we have off-diagonal magnetoelectric susceptibilities, 
  \begin{equation}
    {\rm Im} [\chi^{{\rm me}}_{\beta\alpha}  (\omega)
      + \chi^{{\rm em}}_{\alpha\beta}  (\omega)]\ne0,
    \label{eq:chi-me-off}
  \end{equation}
  the system exhibits NDD for an incident light
  with $E^\omega_\alpha$ and $H^\omega_\beta$  
  for $\alpha,\beta=x,y,z$ ($\alpha\perp\beta$).

  Using the Kubo formula,
  the dynamical susceptibilities $\chi^{\rm mm}_{\gamma \tau} (\omega)$,
  $\chi^{\rm ee}_{\gamma \tau} (\omega)$,  
  $\chi^{\rm me}_{\gamma \tau} (\omega)$,  and  
  $\chi^{\rm em}_{\gamma \tau} (\omega)$
  for $\gamma,\tau=x,y,z$ are described as
  \begin{eqnarray}
    \chi^{\rm mm}_{\gamma \tau} (\omega) & = & 
    \frac{\mu_0}{\hbar N V} 
    \sum_{n} \frac{\langle 0 | \Delta M_\gamma | n \rangle 
      \langle n | \Delta M_\tau | 0 \rangle}
	{\omega_{n0} - \omega -  i\delta}, 
	\label{eq:chi-mm}	\\
    \chi^{\rm ee}_{\gamma \tau} (\omega) & = & 
    \frac{1}{\hbar N V \epsilon_0}
    \sum_{n} \frac{\langle 0 | \Delta P_\gamma | n \rangle 
      \langle n | \Delta P_\tau | 0 \rangle}
	{\omega_{n0} - \omega -  i\delta}, 
    \label{eq:chi-ee} \\
    \chi^{\rm me}_{\gamma \tau} (\omega) & = & 
    \frac{1}{\hbar N V} \sqrt{\frac{\mu_0}{\epsilon_0}}
    \sum_{n} \frac{\langle 0 | \Delta M_\gamma | n \rangle 
      \langle n | \Delta P_\tau | 0 \rangle}
	{\omega_{n0} - \omega -  i\delta},  
	\label{eq:chi-me}	\\
    \chi^{\rm em}_{\gamma \tau} (\omega) & = & 
    \frac{1}{\hbar N V } \sqrt{\frac{\mu_0}{\epsilon_0}}
    \sum_{n} \frac{\langle 0 | \Delta P_\gamma | n \rangle 
      \langle n | \Delta M_\tau | 0 \rangle}
	{\omega_{n0} - \omega -  i\delta}. \quad
    \label{eq:chi-em}	
  \end{eqnarray}
  Here,  $|0\rangle$ is a ground state with the eigenenergy $E_0$,
  $|n\rangle$ is an $n$-th magnetic excitation state with $E_n$, and 
  $\hbar \omega_{n0} = E_n - E_0$.
  $V$ is the unit volume per spin and $N$ is the number of spins.
  Magnetization is defined by
  \begin{eqnarray}
    {\bf M} & = & \sum_i g \mu_B {\bf S}_i,
    \label{eq:M}
  \end{eqnarray}  
  and $\Delta M_\gamma$ is the $\gamma$-component of the
  magnetization  fluctuations from the ground state.
  We only consider the contributions to the electric polarization
  induced by the spin current mechanism,~\cite{katsura05}
  \begin{eqnarray}
    {\bf P} & = & \lambda\sum_{ij} 
     {\bf e}_{ij}
    \times ({\bf S}_i \times {\bf S}_j),
    \label{eq:P_AS}
  \end{eqnarray}
  which exists in generic systems in the presence of spin-orbit couplings.
  Here, ${\bf e}_{ij} \|y$ is the unit vector connecting 
  the nearest-neighbor (n.n.) spins
  ${\bf S}_i$ and  ${\bf S}_j$, and $\lambda$ is a coupling constant
  that is proportional to spin-orbit couplings.
  $\Delta P_\gamma$ in 
  eqs.~(\ref{eq:chi-ee})-(\ref{eq:chi-em}) is the fluctuation of 
  ${\bf P}$.
 
  From eqs.~(\ref{eq:chi-mm}) and (\ref{eq:chi-ee}), we see that
  electromagnon excitation exists 
  if $\langle n | \Delta P_\gamma | 0 \rangle \ne 0$
  for a one-magnon excitation state $|n\rangle$, while a 
  conventional MR occurs for
  $\langle n | \Delta M_\gamma | 0 \rangle \ne 0$.
  Furthermore, from eqs.~(\ref{eq:chi-me-off}), 
  (\ref{eq:chi-me}), and (\ref{eq:chi-em}),
  we have NDD at an electromagnon resonance if
  ${\rm Re} \langle 0 | \Delta M_\tau | n \rangle 
  \langle n | \Delta P_\gamma | 0 \rangle \ne 0$ with $\tau \perp \gamma$
  being satisfied for an excited state $|n\rangle$.
  
\begin{figure}
    \begin{center}
      \includegraphics[width=0.85\columnwidth]{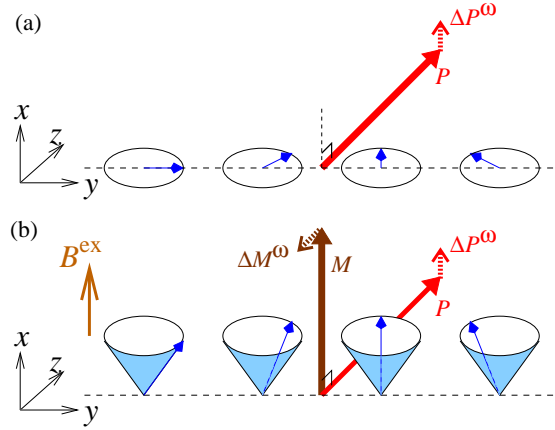}
    \end{center}
    \caption{(Color online) (a) Schematic view of a $yz$-cycloidal state with 
      ${\bf P} \|z$ induced by the spin current mechanism.
      The spin rotation around the $y$-axis accompanies 
      $\Delta {\bf P}^\omega \| x$.
      (b) $yz$-cycloidal conical state with ${\bf P} \|z$ and ${\bf M} \|x$.
      The tilting of the cones around the $y$-axis creates
      $\Delta {\bf P}^\omega \| x$ and $\Delta {\bf M}^\omega \| z$. 
    }
    \label{fig:dynamics}
  \end{figure}

  We consider a $yz$-cycloidal screw magnet, i.e.,
  ${\bf q}_0\| y$ and ${\bf h}\| x$, at ${\bf B}^{\rm ex}=0$, as an example.
  This structure accompanies the electric polarization 
  ${\bf P} \| z$.
  Through the Nambu-Goldstone mode, which modulates the spin structure
  in such a way that the cycloidal plane rotates along the $y$-axis,
  the modulation of the spontaneous electric 
  polarization $\Delta {\bf P}^\omega \| x$ occurs, 
  as shown in Fig.~\ref{fig:dynamics}(a).
  Then, such a mode is activated by
  ${\bf E}^\omega \| x$.\cite{katsura07} 
  Since the modulation of the spin structure
  does not induce a uniform magnetization,
  such an electromagnon mode should not be magneto-active, 
  i.e., $\Delta {\bf M} = 0$.
  Thus, the electromagnon resonance should not accompany NDD.

  On the other hand, 
  under the external magnetic field ${\bf B}^{\rm ex} \| x$,
  spins give a conical structure, as depicted in Fig.~\ref{fig:dynamics}(b).
  Here, we have both the electric polarization ${\bf P} \| z$ and 
  the magnetization ${\bf M} \| x$ in the ground state,
  and thus have a static toroidal moment defined by
  ${\bf T} = {\bf P} \times {\bf M}$~\cite{arima05}.
  If we consider an excited state $| n \rangle$ where
  the conical structure is dynamically tilted 
  along the $y$-axis,~\cite{tewari08}
  both $\Delta{\bf P}^\omega \| x$ and 
  ${\bf \Delta M}^\omega \| z$ exist coherently.
  When the system is excited by an incident light with 
  ${\bf E}^\omega \| x$ and ${\bf H}^\omega \| z$,
  the tilting motion of the conical structure of the spins around
  the $y$-axis exhibits ${\rm Re} \langle 0 | \Delta M_z | n \rangle 
  \langle n | \Delta P_x | 0 \rangle \ne 0$ and thus NDD.
  In a semi-classical picture, NDD is observed 
  at a resonance owing to a magnon with the dynamical toroidal moment 
  ${\bf T}^d = \Delta {\bf P}^\omega \times \Delta {\bf M}^\omega$,
  i.e., 
  a {\it toroidalmagnon}, irrespective of lattice 
  dimensions or symmetries.
  Note that the NDD effect is linearly proportional to 
  $\Delta  P$. Thus, in the weak coupling region,
  it may be detected more easily than
  the electromagnon absorption itself, which is proportional 
  to $(\Delta P)^2$.

  To obtain microscopic details,  
  we investigate a frustrated (quasi-)one-dimensional 
  Heisenberg model,
  \begin{eqnarray}
    {\cal H} & = & J_1 \sum_{\rm n.n.} {\bf S}_i \cdot {\bf S}_j
    + J_2 \sum_{\rm n.n.n.} {\bf S}_i \cdot {\bf S}_j,
    \label{eq:model}
  \end{eqnarray}
  as a canonical spin model for helimagnets. 
  Here, ${\bf S}_i$ is a spin operator on the $i$-th site.
  We assume the ferromagnetic n.n. spin exchange interaction
  $J_1$ and antiferromagnetic
  next-nearest-neighbor (n.n.n.) interaction 
  $J_2$, i.e., $J_1 < 0$ and  $J_2 > 0$.
  In the absence of the external magnetic field $B^{\rm ex}$, 
  the helical spin state  
  is the ground state in the
  parameter range $J_2/|J_1| > 0.25$ for classical spins
  owing to the frustration.
  Hereafter, we consider ${\bf q}_0 \| y$ as the propagation vector, i.e.,
  the helical spin structure along the $y$-direction has 
  a $2 \pi/q_0$ period, where $\cos q_0 = -J_1/ 4 J_2$.
  When ${\bf B}^{\rm ex} $ is applied, a conical spin state emerges.
  The periodicity of the helical structure does not depend 
  on ${\bf B}^{\rm ex} $~\cite{cooper62}.
  We apply the lowest order spinwave expansion from the ground state.

  First, we consider a helical magnet with ${\bf B}^{\rm ex} \perp {\bf q}_0$.
  For ${\bf B}^{\rm ex} \| x$, we have $yz$-cycloidal  
  and conical spin structures 
  with the helicity vector ${\bf h} = {\bf S}_i \times {\bf S}_{i+1} \| x$.
  The ground state spin structure is described as
  \begin{equation}
    {\bf S}_i = (S \cos \theta, \,
    S \sin\theta \sin ({ q}_0  { r}_i), \, 
    S \sin\theta \cos ({ q}_0  { r}_i)),
    \label{eq:g-spin}
  \end{equation}
  where the cone angle $\theta$ is defined as
  \begin{equation}
    2 S \cos \theta = \frac{g \mu_B B^{\rm ex}_x}
    {J_1 (1 - \cos q_0) + J_2 (1 - \cos 2 q_0) }. 
  \end{equation}
  Note that $\theta\to\pi/2$ for $B^{\rm ex}_x\to0$ 
  such that we recover a pure helical structure.
  We have ${\bf P}\| z $ and ${\bf M}\| x$ in the ground state.
  
  Following the procedure indicated in ref.~\citen{cooper62},
  we use the linear spin wave theory in eq.~(\ref{eq:model}).
  We define the local coordinates where the $\xi$-direction
  is aligned to the equilibrium spin directions:
  \begin{eqnarray}
    S^x_i  &=&   -S_i^\zeta \sin \theta + S_i^\xi \cos \theta, \\
    S^y_i  &=&  (S_i^\zeta \cos \theta
    + S_i^\xi \sin \theta ) \sin ({ q}_0 { r}_i)
    + S_i^\eta \cos ({q}_0 { r}_i),\\
    S^z_i  &=&  (S_i^\zeta \cos \theta 
    + S_i^\xi \sin \theta )\cos ({q}_0 { r}_i) 
    - S_i^\eta \sin ({q}_0  { r}_i). \qquad
  \end{eqnarray}
  Hereafter, we choose $q_0 > 0$ for the ground state.
  Then, the Hamiltonian (\ref{eq:model}) can be written in terms of
  local $\zeta$, $\eta$, and $\xi$ coordinates by using $S_i^+, S_i^-$, 
  and $S_i^\xi$.   
  For spin operators, we consider the Holstein-Primakoff approximations:
  \begin{equation}
    S_i^+ = \sqrt{2 S} a_i,\quad S_i^- = \sqrt{2 S} a_i^\dagger, \quad
    S_i^\xi = S - a_i^\dagger a_i.
    \label{eq:HP}
  \end{equation}
  By the Fourier transformations of $a_i$ and $a_i^\dagger$,
  the linear spinwave Hamiltonian in the $q$ space is written as 
  \begin{equation}
    {\cal H} = E_0 + S\sum_q \left[ 2 A_q a_q^\dagger a_q
    + B_q (a_q^\dagger a_{-q}^\dagger + a_q a_{-q}) \right],
    \label{eq:spin_H}
  \end{equation}
  where
  \begin{eqnarray}
    A_q & = & 
       - J_{1} \{ \cos q_0
    + \cos \theta \sin q_0 \sin q
    \nonumber \\ && 
    -  \frac{1}{2} [ (1 + \cos^2 \theta) 
    \cos q_0 + \sin^2 \theta ] \cos q \}
    \nonumber \\ 
    && - J_{2} \{ \cos 2 q_0
    +\cos \theta \sin 2 q_0 \sin 2 q
    \nonumber \\  &&
    - \frac{1}{2} [ (1 + \cos^2 \theta) 
    \cos 2 q_0 + \sin^2 \theta ] \cos 2 q\},
    \nonumber \\    
    B_q & = & 
    \frac{J_{1}}{2} \sin^2 \theta (1 - \cos q_0)  \cos q 
    \nonumber \\  && 
    + \frac{J_{2}}{2} \sin^2 \theta (1 - \cos 2 q_0)  \cos 2 q.
  \end{eqnarray}
  Here, the magnon wave vector $q$ is defined in
  an extended Brillouin zone to clarify distinctions for
  the magnon modes.  The Hamiltonian (\ref{eq:spin_H}) can be diagonalized
  by the standard Bogoliubov transformation,
  \begin{equation}
    \alpha^\dagger_q \equiv c_q a^\dagger_q + s_q a_{-q}, \,\,\,
    \alpha_q \equiv s_q a^\dagger_{-q} + c_q a_q,
    \label{eq:Bogoliubov}
  \end{equation}
  where the coefficients $c_q$ ($c_q = c_{-q}$) 
  and $s_q$ ($s_q = s_{-q}$) are given by
  \begin{eqnarray}
    c_q  =  \frac{2 B_q}{\sqrt{4 B_q^2 -C_q^2 }}, \quad
    s_q  =  \frac{C_q}{\sqrt{4 B_q^2 -C_q ^2}},
  \end{eqnarray}
  for $C_q=A_q + A_{-q} - [(A_q+A_{-q})^2 - 4 B_q^2]^{1/2}$.
  We obtain
  \begin{equation}
    {\cal H} = \sum_q \hbar \omega_q \alpha_q^{\dagger} \alpha_q
    + {\rm const},
  \end{equation}
  where the spin wave frequencies are 
  \begin{eqnarray}
    \hbar \omega_q = S \left[
    (A_q - A_{-q}) +  \sqrt{(A_q + A_{-q})^2 - 4 B_q^2}\,
     \right].
    \label{eq:magnon-omega}
  \end{eqnarray}
  For the helical spin state,
  i.e., $B^{\rm ex} = 0$, $A_q = A_{-q}$ so that
  the magnons at $q$ and $-q$ are degenerate.
  On the other hand, the external magnetic field lifts
  the degeneracy owing to $A_q \neq A_{-q}$, i.e.,
  $\hbar \omega_{q} \neq \hbar \omega_{-q}$ for $B^{\rm ex} \ne 0$.
  
  When the spinwave expansion of ${\bf P}$ contains  
  linear terms of magnon creation and annihilation
  operators $\alpha^\dagger_q$ and $\alpha_q$, 
  the electric component of light creates a one-electromagnon excited state, 
  through the perturbation term ${\cal H}' = -{\rm E}^\omega \Delta P$.
  In the present model, 
  we apply eqs.~(\ref{eq:HP}) and (\ref{eq:Bogoliubov}) 
  to ${\bf P}$ and ${\bf M}$.
  Both the magnetization (\ref{eq:M})
  and the electric polarization  (\ref{eq:P_AS}) can be
  described as
  \begin{eqnarray}
    \Delta M_{\tau} & \sim & \sum_q
    g \mu_B \sqrt{2 SN} \left[ f_\tau^{\rm m} ({ q}) \alpha_q^\dagger 
    + f_\tau^{{\rm m}\,*} ({ q}) \alpha_q \right],\qquad
    \label{eq:M_alpha} \\
    \Delta P_\tau  & \sim & \sum_{q} 
    \lambda S \sqrt{2 S N}
    \left[ f^{\rm e}_\tau ({ q}) \alpha^\dagger_q
    + f^{{\rm e}\,*}_\tau ({ q}) \alpha_q \right].
    \label{eq:P_alpha}
  \end{eqnarray}
  Non-zero coefficients for the the lowest order spinwave expansion
  $ f^{\alpha}_\tau ({ q})$ for $\alpha={\rm e},{\rm m}$ and $\tau=x,y,z$
  are summarized in Table~\ref{tab:fq_omegaq}.
  We only have non-zero contributions from $q=\pm q_0$ magnons,
  which correspond to the electromagnons discussed in ref.~\citen{katsura07}.
  Note that, within the magnetic Brillouin zone,
  $q=\pm q_0$ locates at the $\Gamma$ point.
  At $B^{\rm ex} = 0$, these $q = \pm q_0$ Nambu-Goldstone modes are gapless
  and thus do not show resonance in reality.
  However, under $B^{\rm ex} \neq 0$, the ${ q} = \pm q_0$ 
  magnon excitation has the gap $\hbar \omega_{\pm q_0}$ 
  due to Zeeman energy.
  Thus, the magnetoelectric resonance of the magnon can
  be observed in a finite frequency range where
  a conventional MR is observed.

  \begin{table}
    \begin{center}
      \caption{Non-zero coefficients $f_\tau^{\alpha} ({ q})$  
	for $\tau=x,y,z$ and $\alpha=\mbox{m,e}$
	in a $yz$-cycloidal conical spin structure 
	with ${\bf q_0} \| y$ and ${{\bf B}^{\rm ex} \|\bf h}\| x$.
     }
      \label{tab:fq_omegaq}
      \begin{tabular}{cl}
	$\Delta M_y$ & $f^{\rm m}_{y} ({\pm q_0}) 
	= i [ (1 \pm \cos\theta) c_{q_0} 
	+ (1 \mp \cos\theta) s_{q_0} ]/4$ \\
	$\Delta M_z$ & $f^{\rm m}_{z} ({\pm q_0}) = \pm 
	[ (1 \pm \cos\theta) c_{q_0} 
	+ (1 \mp \cos\theta) s_{q_0} ]/4$ \\
	\hline
	$\Delta P_x$ &  $f^{\rm e}_{x} ({\pm q_0}) = -
	(c_{q_0} - s_{q_0}) \sin^2 \theta \, \sin q_0/2$ \\
      \end{tabular}
    \end{center}
  \end{table}
  
  Within a linear spinwave treatment, the susceptibilities
  $\chi^{\rm mm}_{\gamma\tau} (\omega)$,  
  $\chi^{\rm ee}_{\gamma\tau} (\omega)$,  
  $\chi^{\rm me}_{\gamma\tau} (\omega)$, 
  and $\chi^{\rm em}_{\gamma\tau} (\omega)$    
  are given as
  \begin{equation}
    \chi^{\alpha\beta}_{\gamma\tau} (\omega) 
    = \frac{C^{\alpha\beta}}{\hbar V}
    \sum_q \frac{f^{\alpha *}_\gamma ({ q}) f^{\beta}_\tau ({ q})}
    {\omega_{q} - \omega - i \delta} 
    \,\,\,\,\,\,\left(
    \begin{array}{c}
      \alpha,\beta={\rm e},{\rm m}\\
      \gamma,\tau=x,y,z
    \end{array} \right),
    \label{eq:chi_gen}
  \end{equation}
  where $C^{\rm mm} = 2 S (g \mu_B)^2 \mu_0$,
  $C^{\rm ee} = 2 S^3 \lambda^2/\epsilon_0$,
  and 
  $C^{\rm me} = C^{\rm em} = 2 S^2 \lambda g \mu_B \sqrt{\mu_0/\epsilon_0}$.
  Namely, when $|f^{\rm m}_\tau(q)|^2 \ne 0$, 
  the resonance due to the magnon 
  excitation at $\omega=\omega_q$
  is induced by $H^\omega_\tau$,
  while  $|f^{\rm e}_\gamma(q)|^2 \ne 0$ gives an electromagnon 
  excited by $E^\omega_\gamma$.
  For a non degenerate magnon
  ($\omega_q \neq \omega_{-q}$),
  ${\rm Re} \, [ f^{\rm m\,*}_\tau(q) f^{\rm e}_\gamma(q)] \ne 0$ 
  ($\tau \perp \gamma$)
  ensures NDD for electromagnetic waves 
  with $H^\omega_\tau$ and $E^\omega_\gamma$
  with an intensity  $\propto C^{\rm me}$.

  Since $|f^{\rm e}_x (\pm q_0)|^2\ne 0$ as shown in Table \ref{tab:fq_omegaq},
  these magnons give electric resonances 
  for $\chi^{\rm ee}_{xx} (\omega)$ 
  at $\omega = \omega_{q_0}$ and $\omega_{-q_0}$.
  This indeed indicates that MRs can be 
  driven by ${\bf E}^\omega \| x$ as absorptions by electromagnons.
  We also see $|f^{\rm m}_y (\pm q_0)|^2=|f^{\rm m}_z (\pm q_0)|^2\ne 0$; 
  thus, ${\bf H}^\omega \perp x$ induces MR
  at the ${\bf q} = \pm q_0$ magnon energy~\cite{cooper62}.
  These modes show NDD for an incident light with ${\bf E}^\omega \| x$,
  ${\bf H}^\omega \| z$, and ${\bf k} \|y$, since
  $\chi^{{\rm me}}_{zx}  (\omega_{\pm q_0}) 
  = \chi^{{\rm em}}_{xz} (\omega_{\pm q_0})
  \propto f_z^{{\rm m} *}(\pm q_0) f_x^{\rm e}(\pm q_0)
  + f_x^{{\rm e} *}(\pm q_0)  f_z^{\rm m} (\pm q_0) \ne 0$.
  In this manner, the electromagnon at $q = \pm q_0$ has a
  dynamical toroidal moment ${\bf T}^d \neq 0$ and thus 
  is identified as a toroidalmagnon.
  At $B^{\rm ex} = 0$, however, 
  the $\pm q_0$ magnons become degenerate; thus,
  the cross-correlated effects 
  at the $\omega = \omega_{q_0} $ and $\omega_{-q_0}$ peaks
  merge to cancel themselves:
  $\chi^{{\rm me}}_{zx}  (\omega_{q_0}) 
  \propto  f_z^{{\rm m} *}(q_0) f_x^{\rm e}(q_0) 
  + f_z^{{\rm m} *} (-q_0) f_x^{\rm e}(-q_0) = 0$.
  
  Other components of susceptibilities are 
  also obtained from Table \ref{tab:fq_omegaq}.
  Magnons at $q= \pm q_0$ 
  contribute to the off-diagonal magnetic susceptibility
  $\chi^{\rm mm}_{yz} (\omega) = -\chi^{\rm mm}_{zy} (\omega)$ 
  at $B^{\rm ex}\ne0$,
  which give a conventional Faraday rotation.  
  For magnetoelectric susceptibilities,
  $\chi^{{\rm me}}_{yx} (\omega)$ and 
  $\chi^{{\rm em}}_{xy} (\omega)$ are also non-zero.
  However, they do not cause NDD since 
  ${\rm Im} [ \chi^{{\rm me}}_{yx} (\omega) 
  + \chi^{{\rm em}}_{xy} (\omega)] =0$.~\cite{memo}

  By reversing the direction of the external magnetic fields 
  $B^{\rm ex} \rightarrow -B^{\rm ex}$, 
  $\chi^{{\rm me}}_{zx}  (\omega_{\pm q_0}) 
  \rightarrow -\chi^{{\rm me}}_{zx}  (\omega_{\mp q_0})$
  is obtained owing to $\theta \rightarrow \pi - \theta$ 
  and $\hbar \omega_{\pm q_0} \rightarrow \hbar \omega_{\mp q_0}$.
  Thus, reversing the light propagation direction is equivalent to
  reversing the external magnetic field direction 
  for the absorption process.

\begin{figure}
    \begin{center}
      \includegraphics[width=0.85\columnwidth]{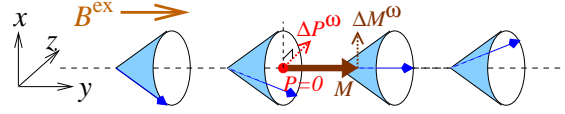}
    \end{center}
    \caption{(Color online) Schematic view of a
      $xz$-proper conical state with  ${\bf M} \|y$ but ${\bf P} =0$.
      The tilting of the cones around the $z$-axis creates
      $\Delta {\bf P}^\omega \| z$ and $\Delta {\bf M}^\omega \| x$.}
    \label{fig:proper}
  \end{figure}

  Next, we investigate the case ${\bf B^{\rm ex}} \| {\bf q}_0 \|y$,
  where 
  a proper conical spin state is stabilized, as shown
  in Fig.~\ref{fig:proper}.
  Note that the ground state does not have a spontaneous electric 
  polarization
  owing to ${\bf e}_{ij} \| {\bf h} \|y$
  such that ${\bf T} = {\bf P} \times {\bf M} = 0$ statically.
  If we consider an excitation state where helical structures are uniformly
  tilted along the $z$-axis, i.e., ${\bf h}$ is rotated
  around the $z$-axis to create $\Delta{\bf h} \| x$, the 
  electric polarization 
  $\Delta{\bf P}^\omega \sim ({\bf e}_{ij} \times \Delta {\bf h} )\| z$ 
  is induced,
  which couples to the external electric field ${\bf E}^{\omega} \| z$.
  Furthermore, in the case of ${\bf B}^{\rm ex} \ne 0$,  
  $\Delta{\bf M}^\omega \| x$ is also seen in this excited state.
  Thus,
  the mode is accompanied by
  the dynamical toroidal moment ${\bf T}^d \ne 0$ even in the absence
  of the static ${\bf T}$, such that
  NDD for an incident light with ${\bf E}^{\omega} \| z$ and
  ${\bf H}^{\omega} \| x$ is expected.
  The tilting of the conical structure along the $x$-axis is symmetry-wise 
  equivalent. This excitation state will induce 
  $\Delta{\bf P}^{\omega} \| x$ and $\Delta{\bf M}^{\omega} \| z$, 
  and NDD for an incident light with 
  ${\bf E}^{\omega} \| x$ and
  ${\bf H}^{\omega} \| z$ is similarly expected.
  
  For details, we perform a similar calculation for the
  spinwave expansions in a proper conical spin state
  with ${\bf q_0} \|{{\bf B}^{\rm ex} \|\bf h}\| y$.
  We obtain $\Delta {\bf P}$ and $\Delta {\bf M}$
  within the linear combinations of spinwave operators, 
  as listed in Table \ref{tab:fq_omegaq_pro}.
  We see that magnons with $q=\pm q_0$ are electro-active,
  owing to $|f^{\rm e}_x(\pm q_0)|^2 = |f^{\rm e}_z(\pm q_0)|^2 \ne 0$.
  Namely, these modes are electromagnons that couple 
  to ${\bf E}^\omega \perp y$, irrespective of ${\bf B}^{\rm ex}$.
  Moreover, they are also magneto-active,
  since $|f^{\rm m}_x(\pm q_0)|^2 = |f^{\rm m}_z(\pm q_0)|^2 \ne 0$
  at ${\bf B}^{\rm ex} \ne 0$.
 
  \begin{table}
    \begin{center}
      \caption{Non-zero coefficients $f_\tau^{\alpha} ({ q})$  
	for $\tau=x,y,z$ and $\alpha=\mbox{m,e}$
	in an $xz$-proper conical spin structure 
	with ${\bf q_0} \|{{\bf B}^{\rm ex} \|\bf h}\| y$.
      }
      \label{tab:fq_omegaq_pro}
      \begin{tabular}{cl}
	$\Delta M_x$ & $f_x^{\rm m} (\pm q_0)= -i 
	[ (1 \pm \cos\theta) c_{q_0} 
	+ (1 \mp \cos\theta) s_{q_0} ]/4$  \\
	$\Delta M_z$ & $f_z^{\rm m} (\pm q_0)= 
	\pm [ (1 \pm \cos\theta) c_{q_0} 
	+ (1 \mp \cos\theta) s_{q_0} ] /4$  \\
	\hline
	$\Delta P_x$ & 
	$f_x^{\rm e} (\pm q_0)= -(c_{q_0} - s_{q_0}) 
	\sin^2 \theta \, \sin q_0/2$  \\
	$\Delta P_z$ &  $f_z^{\rm e} (\pm q_0)= \mp i 
	(c_{q_0} - s_{q_0}) \sin^2 \theta \, \sin q_0/2$ 
      \end{tabular}
    \end{center}
  \end{table}

  Let us study NDD in this configuration in detail.
  For  ${\bf E}^{\omega} \| z$ and
  ${\bf H}^{\omega} \| x$, we have
  $ {\rm Im} [\chi^{{\rm me}}_{xz} (\omega_{\pm q_0})
    +\chi^{{\rm em}}_{zx} (\omega_{\pm q_0})]\ne 0$, 
  owing to $ f_x^{\rm m*} (\pm q_0) f_z^{\rm e}(\pm q_0) 
  =f_z^{\rm e*}(\pm q_0) f_x^{\rm m} (\pm q_0) \ne 0$. 
  Similarly, we have
  ${\rm Im} [\chi^{{\rm me}}_{zx} (\omega_{\pm q_0})
    +\chi^{{\rm em}}_{xz} (\omega_{\pm q_0})]\ne 0$  
  for  ${\bf E}^{\omega} \| x$ and
  ${\bf H}^{\omega} \| z$.  At $B^{\rm ex} = 0$, however, 
  the cross-correlated effects caused by $\pm q_0$-magnons
  are canceled owing to their degeneracy. 
  Namely, for an incident light with ${\bf k}\|y$, 
  we see NDD for any polarizations
  of ${\bf E}^{\omega} $ and  ${\bf H}^{\omega}$, 
  in the presence of $B^{\rm ex} \ne 0$.
  Note that this is the Faraday configuration ${\bf k} \| {\bf M}$ 
  such that
  circularly polarized modes are the eigenmodes of the Maxwell equation
  in the presence of rotational symmetry along ${\bf k} \| {\bf M}$;
  thus, the so-called magnetochiral dichroism is observed. 
  Details will be reported elsewhere.
  As in the cycloidal cone case, 
  reversing the light propagation direction is equivalent to
  reversing the external magnetic field direction 
  for the absorption process:
  $B^{\rm ex} \rightarrow -B^{\rm ex}$
  ($\theta \rightarrow \pi - \theta$ 
  and $\hbar \omega_{\pm q_0} \rightarrow \hbar \omega_{\mp q_0}$)
  induces $\chi^{{\rm me}}_{zx}  (\omega_{\pm q_0}) 
  \rightarrow -\chi^{{\rm me}}_{zx}  (\omega_{\mp q_0})$
  and $\chi^{{\rm me}}_{xz}  (\omega_{\pm q_0}) 
  \rightarrow -\chi^{{\rm me}}_{xz}  (\omega_{\mp q_0})$.

  Microscopically, NDD arises 
  owing to the off-diagonal matrix elements of 
  both the magnetic  and  electric dipole operators 
  ${\rm Re} \langle 0 | \Delta M_\gamma |n \rangle 
  \langle n| \Delta P_\tau | 0 \rangle$ ($\gamma \ne \tau$) 
  [see eqs.~(\ref{eq:chi-me-off}), (\ref{eq:chi-me}), and (\ref{eq:chi-em})].
  In general, the magnetic dipole operator is even (odd) under
  space-inversion $I$ (time-reversal $R$) symmetry operation, whereas
  the electric dipole operator is odd (even) under $I$ ($R$).
  Thus, in magnetic structures with $I$ and $R$ symmetry invariances, 
  the real parts of the off-diagonal 
  matrix elements are zero, i.e, NDD does not arise.
  In fact, the matrix elements
  in pure helical structures vanish owing to the $R$ symmetry invariance,
  and NDD is induced only in conical structures
  where $I$ and $R$ symmetries are broken.

  To summarize, we investigated 
  NDD that arises in conical spin structures via the spin-current mechanism. 
  NDD is generally induced by the excitation of 
  a magnon accompanied by a dynamical toroidal moment, i.e,
  toroidalmagnon. 
  In a conical structure, the fluctuation of the dynamical toroidal moment
  is observed with cone oscillations, as shown
  in Figs.~\ref{fig:dynamics} and \ref{fig:proper}. 
  Such modes are adiabatically connected to the $q = \pm q_0$
  Nambu-Goldstone modes,
  and thus we conclude that a toroidalmagnon should ubiquitously exist 
  in conical spin structures.
  Let us finally note that the observation of NDD 
  via the magnon excitation in a perovskite helimagnet
  has been reported in ref.~\citen{takahashi11} 
  quite recently, 
  which is qualitatively explained by our theory.

  \section*{Acknowledgment}

  We thank I.~K\'{e}zsm\'arki, S.~Bord\'acs, Y.~Takahashi, N.~Kida, 
  T.~Arima, R.~Shimano, N.~Nagaosa,  and Y.~Tokura for fruitful discussion.
  This work is supported in part by Grants-In-Aid for Scientific 
  Research from the Ministry of Education, Culture, 
  Sports, Science and Technology (MEXT), Japan.


\begin{thebibliography}{10}
\bibitem{memo}
  In ref.~\citen{cano09}, 
  NDD for the non conical cycloidal spin structure due
  to $\chi^{\rm me}_{yx} (\omega_{\pm q_0})$ is discussed. However, in reality,
  the $\chi^{\rm me}_{yx} (\omega_{\pm q_0})$ term does not induce NDD owing to
  $\chi^{\rm me}_{yx} (\omega_{\pm q_0}) = -\chi^{\rm em}_{xy} (\omega_{\pm
  q_0})$ as a consequence of a time-reversal symmetry.
\end{thebibliography}
\end{document}